\documentclass[a4paper]{spie}  
\usepackage[]{graphicx}
 
\newcommand{\um}{$\mu$m}
\newcommand{\degree}{$^{\circ}$}

\title{The GRAVITY Coud\'{e} Infrared Adaptive Optics (CIAO) system for the VLT Interferometer} 

\author{S. Kendrew\supit{a}, S. Hippler\supit{a}, W. Brandner\supit{a}, Y. Cl\'{e}net\supit{b}, C. Deen\supit{a}, E. Gendron\supit{b}, A. Huber\supit{a}, R. Klein\supit{a}, W. Laun\supit{a}, R. Lenzen\supit{a}, V. Naranjo\supit{a}, U. Neumann\supit{a}, J. Ramos\supit{a}, R.-R. Rohloff\supit{a}, P. Yang\supit{a}, F. Eisenhauer\supit{c}, E. Fedrigo\supit{d}, M. Suarez-Valles\supit{d}, A. Amorim\supit{e}, K. Perraut\supit{f}, G. Perrin\supit{b} and C. Straubmeier\supit{g}
\skiplinehalf
\supit{a}Max Planck Institute for Astronomy, K\"{o}nigstuhl 17, 69117 Heidelberg, Germany; \\
\supit{b}LESIA-Observatoire de Paris, CNRS UMR 8109, UPMC Univ. Paris 6, Université Paris Diderot, 5 place Jules Janssen, 92195, Meudon, France;\\
\supit{c}Max Planck Institute for extraterrestrial Physics, PO Box 1312, Giessenbachstr.,  85748 Garching, Germany; \\
\supit{d}ESO, Karl-Schwarzschild-Str. 2, 85748 Garching, Germany; \\
\supit{e}Laboratório de Sistemas, Instrumentação e Modelação em Ciências e Tecnologias do Ambiente e do Espaço (SIM), Lisbon and Porto, Portugal; \\
\supit{f}UJF-Grenoble 1/CNRS-INSU, Institut de Plan\'{e}tologie et d'Astrophysique de Grenoble, France; \\
\supit{g}Physikalisches Institut, University of Cologne, Germany;\\
}

\authorinfo{Send correspondence to S. Kendrew, E-mail: kendrew@mpia.de}

\begin{document} 
  \maketitle 

\begin{abstract}
GRAVITY is a second generation instrument for the VLT Interferometer, designed to enhance the near-infrared astrometric and spectro-imaging capabilities of VLTI. Combining beams from four telescopes, GRAVITY will provide an astrometric precision of order 10 micro-arcseconds, imaging resolution of 4 milli-arcseconds, and low and medium resolution spectro-interferometry, pushing its performance far beyond current infrared interferometric capabilities. To maximise the performance of GRAVITY, adaptive optics correction will be implemented at each of the VLT Unit Telescopes to correct for the effects of atmospheric turbulence. To achieve this, the GRAVITY project includes a development programme for four new wavefront sensors (WFS) and NIR-optimized real time control system. These devices will enable closed-loop adaptive correction at the four Unit Telescopes in the range 1.4-2.4~$\mu$m. This is crucially important for an efficient adaptive optics implementation in regions where optically bright references sources are scarce, such as the Galactic Centre. We present here the design of the GRAVITY wavefront sensors and give an overview of the expected adaptive optics performance under typical observing conditions. Benefiting from newly developed SELEX/ESO SAPHIRA electron avalanche photodiode (eAPD) detectors providing fast readout with low noise in the near-infrared, the AO systems are expected to achieve residual wavefront errors of $\leq$400 nm at an operating frequency of 500 Hz.

\end{abstract}



\section{INTRODUCTION}
\label{sec:intro}  

GRAVITY\cite{grav_messenger} is a second generation 4-beam combiner instrument for ESO's Very Large Telescope Interferometer (VLTI) at Paranal Observatory. Operating in the near-infrared (NIR) K-band (2.2~\um), it will provide of order 10-micro-arcsec astrometric precision and phase-referenced imaging with 4 milli-arcsec resolution. With fibre-fed integrated optics, wavefront sensors, fringe tracker, beam stabilisation and a novel metrology concept, GRAVITY's performance will bring a dramatic improvement on current NIR interferometric instrumentation in sensitivity as well as astrometric precision. First astronomical light at VLTI is planned for 2014.

Key science goals for GRAVITY include:
\begin{itemize}
	\item{Probing the physics in the immediate vicinity of the Galactic Centre black hole\cite{paumard11, vincent11};}
	\item{Detecting and measuring the masses of black holes in massive star clusters;}
	\item{Study the details of mass accretion and jets in young stellar objects; and }
	\item{Characterize the detailed motions of exoplanets, binary stars and young stellar discs.}
\end{itemize}

Crucial to achieving its ambitious science goals is the implementation of adaptive optics (AO) correction to provide seeing-improved beams to the instrument. The GRAVITY project includes a development programme for 4 new wavefront sensors (WFS) for the VLT Unit Telescopes, which will allow the measurement of the atmospheric turbulence correction in the NIR H and K bands (1.4-2.4~\um). These WFS will be installed in the Coud\'{e} space for each UT, and interface directly with the existing Multiple Application Curvature Adaptive Optics (MACAO) systems at VLT. 

In this paper, we present an overview of the design and development of these WFS units and their real time control, as well as the expected on-sky performance in the various operational modes of the AO system. The following sections will provide basic descriptions of the GRAVITY instrument, its subsystems and operational concept, and the AO architecture for the instrument.

\subsection{The GRAVITY instrument}
\label{sec:gravity_inst}

\begin{figure}
	\centering
	\includegraphics[width=15cm]{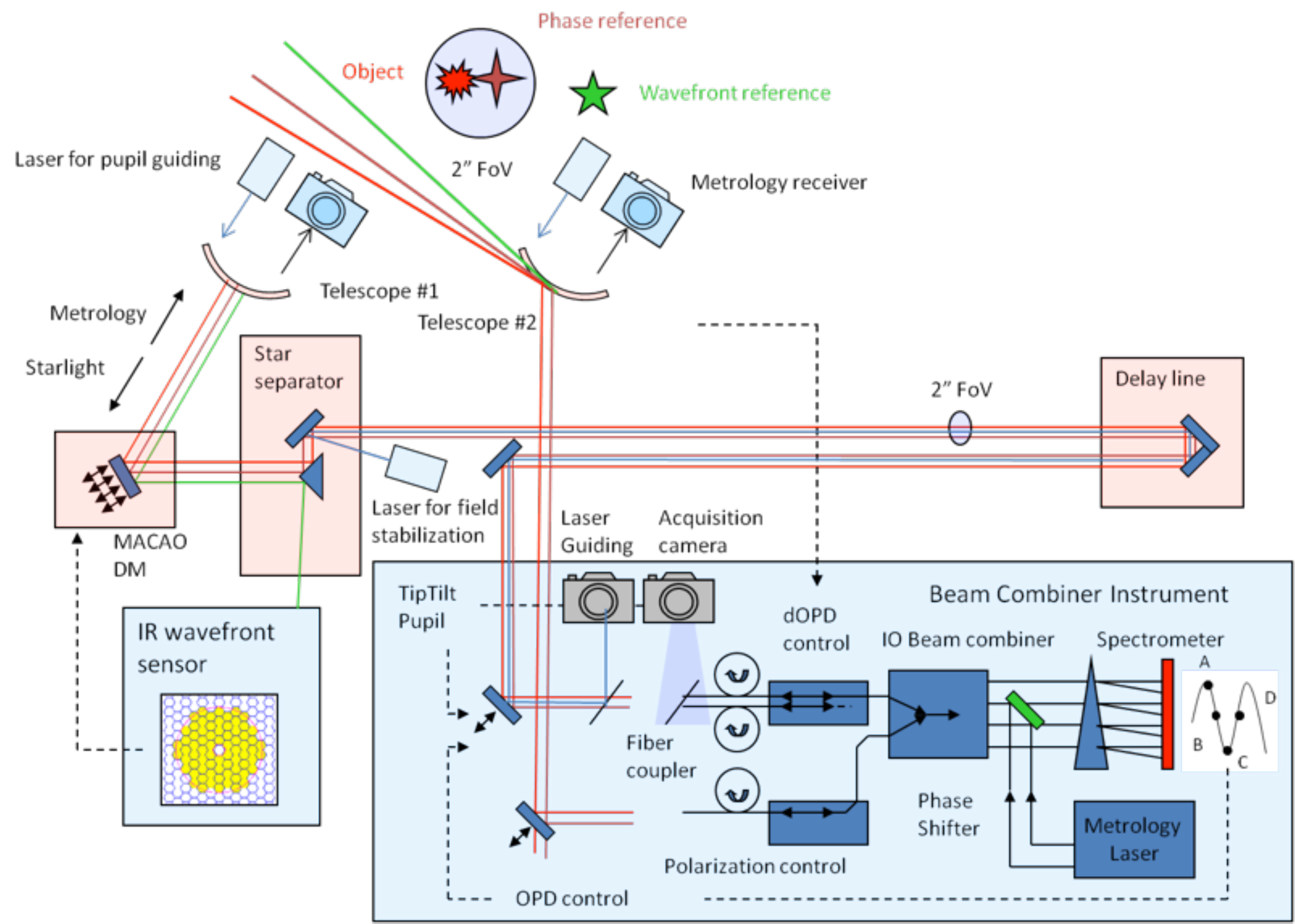}
	\caption{Schematic overview of the GRAVITY instrument, showing the general working principle of the instrument (for the simplified case of 2 input beams).}\label{fig:grav_diagram}
\end{figure}

The GRAVITY instrument consists of three major components: the beam-combiner instrument, the laser metrology system, and the wavefront sensors (see Fig.~\ref{fig:grav_diagram}). In the beam combiner instrument, the four input beams from the Unit or Auxiliary Telescopes are combined to produce interferograms from six baselines simultaneously. Fringe tracking is implemented internally to the instrument to compensate in real time for longitudinal vibrations and atmospheric piston\cite{choquet_spie10}. The input beams are fed via optical fibres into two integrated optics beam combiners\cite{jocou_spie12}.

The fringes of both science and fringe tracking targets are spectrally dispersed by the spectrometers inside the beam combiner cryostat, with the science spectrometer providing spectral resolutions of R$\sim$20, $\sim$500 or $\sim$4000\cite{fischer_spie12, straubmeier_spie10}. The fringe tracking spectrometer provides low-resolution dispersion only. The laser metrology system tracks a reference beam through the observatory, allowing for a precise measurement of the optical path.

\subsection{Adaptive optics for GRAVITY}
\label{sec:gravity_ao}

When observing with the VLT UTs, GRAVITY's AO system will use parts of the existing adaptive optics (AO) hardware to correct for atmospheric turbulence effects. Wavefront sensing will however be performed with dedicated WFS operating at NIR wavelengths, interfacing with the MACAO DMs already present at the UTs. Light is fed to the WFS from the telescopes via the Star Separator Systems (STS)\cite{delplancke_spie04}, installed at the Coud\'{e} foci of each UT as part of the PRIMA instrument, which can pick up 2 stars anywhere inside a 2' field.

The top-level functional requirements for the WFS are:
\begin{itemize}
	\item{provide real-time processing of wavefront distortions;}
	\item{allow for on-axis and off-axis wavefront sensing;}
	\item{provide sensing in the H and K bands (1.45-2.45~$\mu$m);}
	\item{allow use by other VLTI instruments;}
	\item{interface to the existing MACAO deformable mirrors.}
\end{itemize}

In on-axis mode, the science target itself is used as AO reference source, in this case a beam splitter redirects $\sim$50\% of the light to the WFS, with the rest passing to the VLTI tunnels towards the GRAVITY beam combiner. In off-axis mode, where a dedicated reference source is used, the WFS receives all the light from the AO reference.

Performance requirements for the WFS are as follows:
\begin{itemize}
	\item{mK=7: Strehl 25\% for 7'' off-axis AO reference source at 30\degree zenith angle}
	\item{mK=7: Strehl 35\% for on-axis AO reference source at 30\degree zenith angle}
	\item{mK=10: Strehl 10\% on-axis at zenith}
	\item{RMS tip/tilt at the STS in the Coud\'{e} room of $<$ 10 mas on sky for mK = 7, $<$ 20 mas on sky for mK = 10}
	\item{Piston fluctuations of $<$ 25 nm RMS over 48 ms, $<$ 125 nm RMS over 290 ms,  $<$ 2000 nm RMS over 10 min.}
\end{itemize}

\section{TECHNICAL IMPLEMENTATION OF THE GRAVITY AO SYSTEM}
\label{sec:wfs}

The main new hardware under development for the CIAO system is the NIR wavefront sensor. In addition, a new real time control system will be implemented for the operation of the AO system. This section describes the optical and cryo-mechanical design of the full WFS unit, as well as the control architecture and detector characteristics.

The wavefront sensors follow a straightforward Shack-Hartmann design. The incoming light is passed through a lenslet array, with each spot refocused onto the detector, from where the distorted wavefront is characterized. 
\subsection{Wavefront sensor optical design}
\label{sec:wfs_optics}

\begin{figure}
	\centering
	\includegraphics[width=12cm]{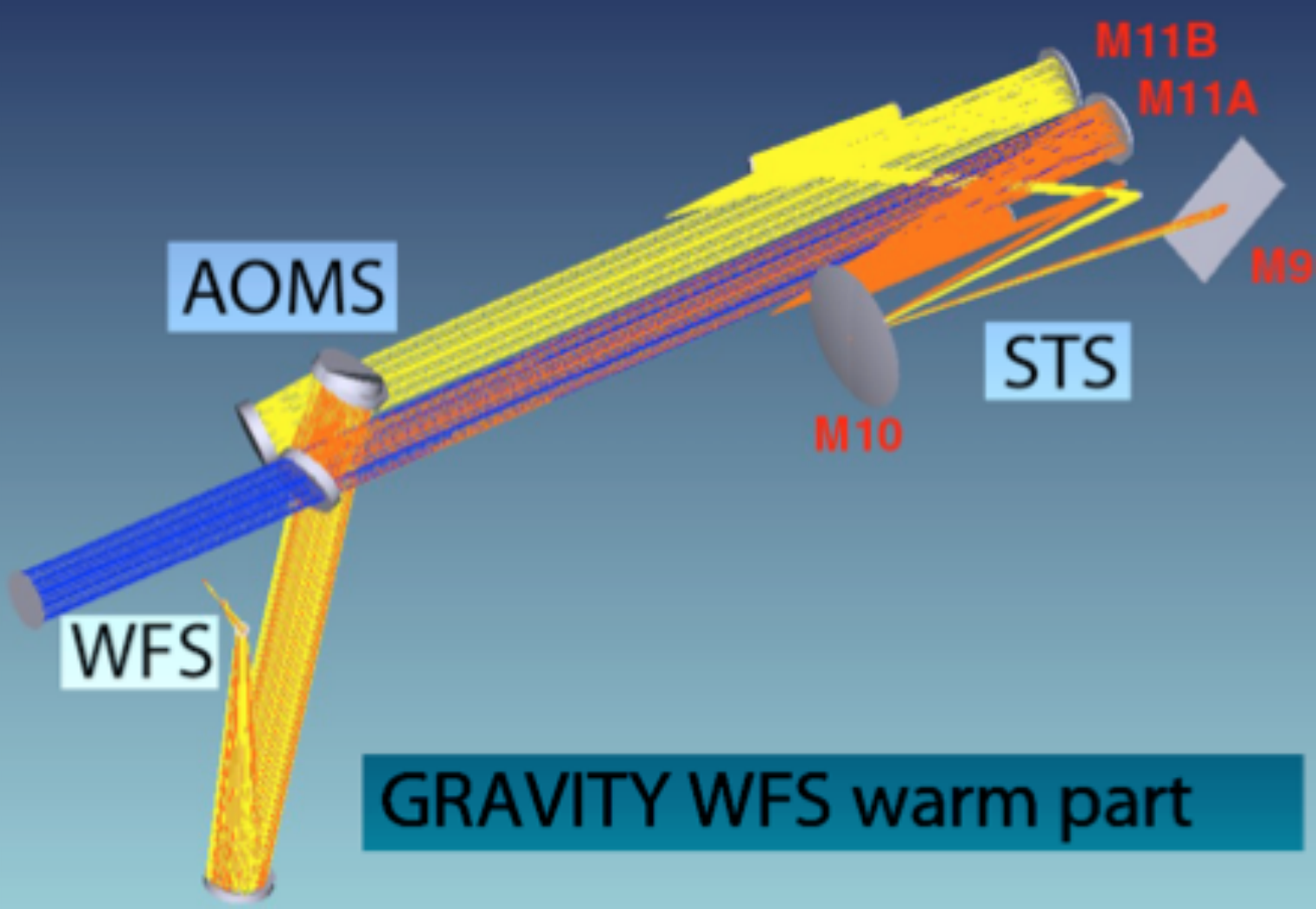}
	\caption{Overview of the WFS warm optics, showing the AO mode selector assembly (AOMS), parabolic mirror, and relay topics to the WFS cryostat. The mirrors labelled M9, M10, M11A/B are part of the Star Separator System, which is not part of the WFS.}\label{fig:wfs_warm}
\end{figure}

Each WFS unit comprises a warm and cold assembly. The warm optics, shown in Fig.~\ref{fig:wfs_warm}, relay the light from the telescope and the STS units to the cryostat. The entrance into the WFS is the AO mode selector consisting of a dichroic and three flat mirrors mounted onto a translation stage, which chooses one of three operational modes for the WFS: (i) on-axis mode, splitting the beam 50/50 between the WFS and the VLTI laboratory; (ii) off-axis mode, 100\% of the light into the WFS, and (iii) visible AO mode, using the existing visible-light MACAO system. 

The light is subsequently folded by parabolic and flat mirrors to the pupil derotator, which maintains pupil orientation throughout an observation. The parabolic mirror will be mounted in a motorized gimbal for optical alignment and focus adjustment. The derotator follows a simple K-mirror design in protected gold-coated aluminium.

\begin{figure}
	\centering
	\includegraphics[width=12cm]{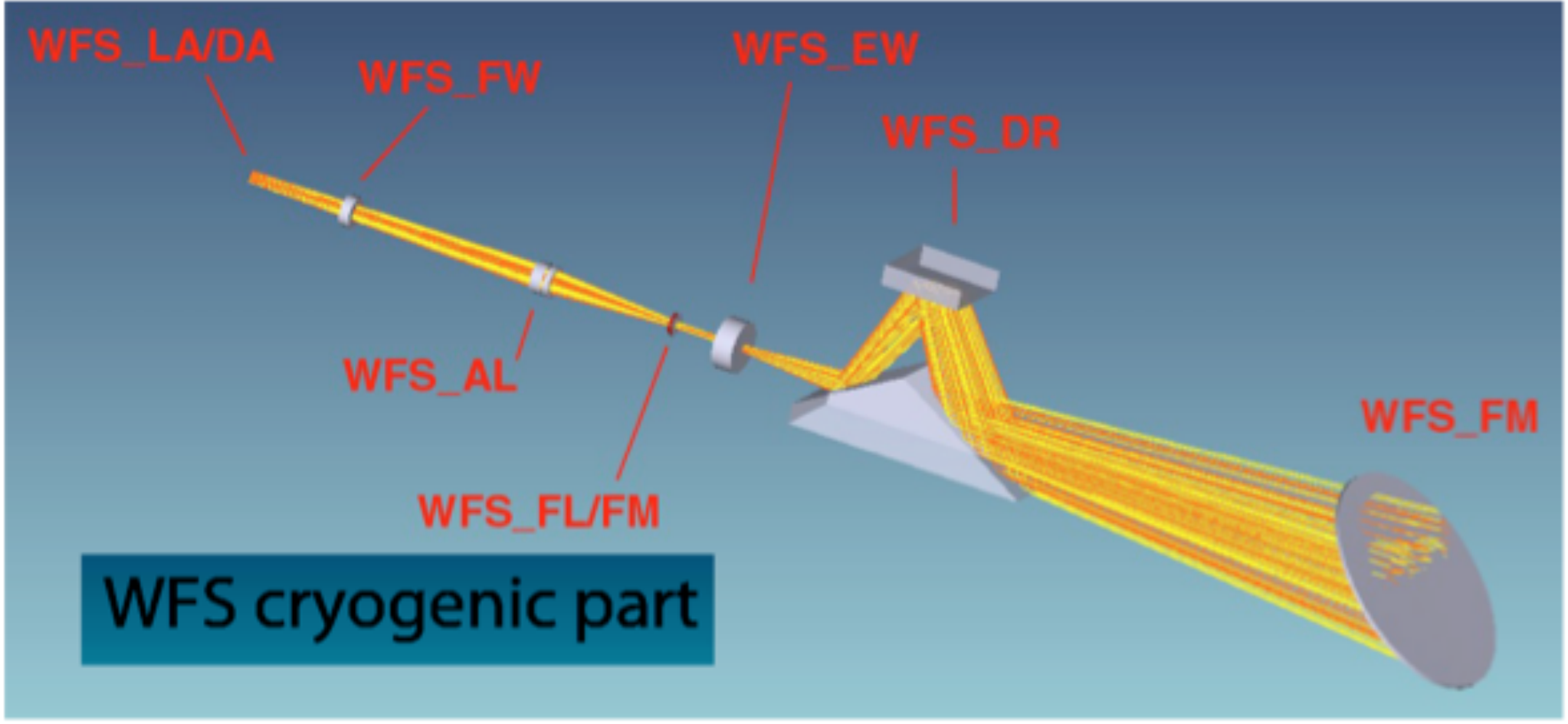}
	\caption{Overview of the light path into the WFS cryostat. A fold mirror (WFS\_FM) directs the light to the pupil derotator (WFS\_DR). The entrance window (WFS\_EW) marks the transition to the cold optics: the field lens (WFS\_FL), achromat (WFS\_AL), filter wheel (WFS\_FW), and finally to the lenslet array and detector (WFS\_LA/DA).}\label{fig:wfs_cold}
\end{figure}

The cold optics are kept at inside a bath cryostat (see Fig.~\ref{fig:wfs_cold}), the cooling concept is described in Section~\ref{sec:wfs_cryomech}. Inside the cryostat, a field lens compensates for pupil motion using a piezoelectric actuator. After this, the light is relayed to the lenslet array via a filter wheel, holding bandpass and neutral density filters. The lenslet array focuses 9 x 9 spots onto the detector from 192~\um~diameter micro-lenslets, each with a focal length of 2.09 mm. 

Further details of the optical design are presented by Yang et al.\cite{yang_spie12} in these proceedings.

\subsection{Detector and read-out electronics}
\label{sec:wfs_detector}

Table~\ref{tab:selex_req} summarises the key requirements for detector performance for the WFS.

The GRAVITY WFS will for the first time use recently developed high-speed IR electron avalanche photodiode (eAPD) arrays, developed by ESO and SELEX-Galileo. The SELEX/ESO SAPHIRA arrays have been shown to reach read noise levels of $<$3 e- and quantum efficiency (QE) of $\sim$40\% at 80K\cite{finger_spie12}. The arrays measure 320 $\times$ 256 24\um~pixels, of which the WFS requires a read-out window of 72 $\times$ 72 pixels for the spot array (9 subapertures, 8 pixels/subaperture). The detector noise characteristics, QE and cosmetics are all significantly improved at lower temperatures, and to benefit from this a cryostat development programme is under way to lower the WFS operating temperature to 40K (see Section~\ref{sec:wfs_cryomech}).

The detector read-out is managed by ESO's Next Generation Controller (NGC)\cite{finger_spie12}, a new high-speed data acquisition system, to be housed inside the Coud\'{e} lab close to the detector array.
                                                                 
\begin{table}
	\centering
	\begin{tabular}{|l|c|c|}
		\hline
		\textbf{Parameter} & \textbf{Goal} & \textbf{Requirement} \\
		\hline
		Spectral range & 1.4-2.4~\um & 1.4-2.4~\um \\
		QE & 70\% & $\geq$50\% \\
		Frame size & 72 $\times$ 72 pix & 72 $\times$ 72 pix \\
		Frame rate & $\geq$500 frames/s & $\geq$500 frames/s \\
		Pixel clock & $\leq$5 MHz & $\leq$5 MHz \\
		Shortest integration time & 2 ms & 2 ms \\
		Read noise & 3 e$^-$/pix (rms, CDS) & 6.5 e$^-$/pix (rms, CDS) \\
		\hline
	\end{tabular}   
	\smallskip
	\caption{Requirements for the performance of the IR detectors for the GRAVITY WFS.}\label{tab:selex_req}
\end{table}

\subsection{Wavefront sensor cryo-mechanical design}
\label{sec:wfs_cryomech}
 
\begin{figure}
	\centering
	\includegraphics[width=13cm]{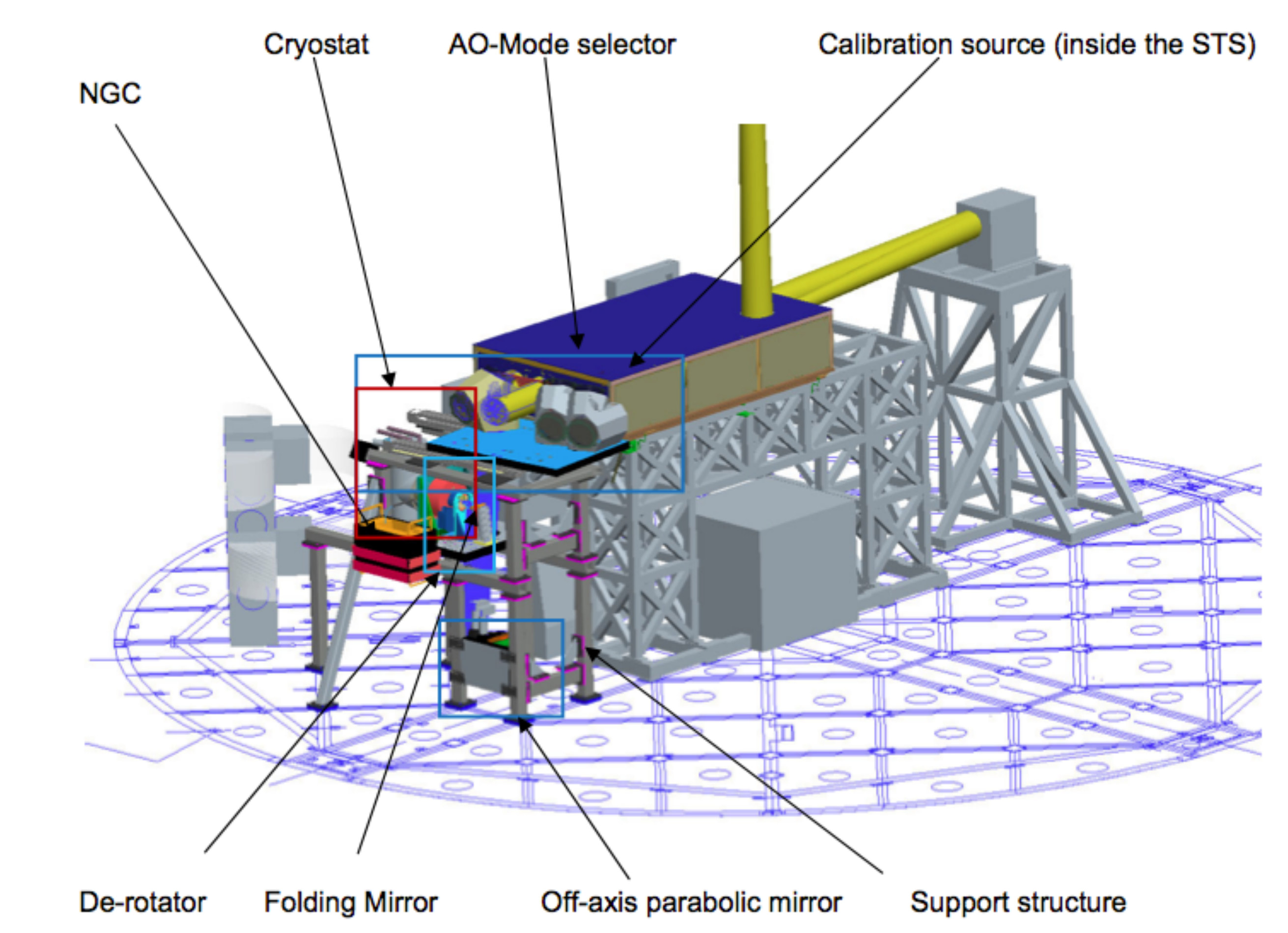}
	\caption{Mechanical overview of the GRAVITY WFS in the VLT UT Coud\'{e} lab, showing the incoming light beam from the telescope (yellow) and the Star Separator System (STS), which forms the optical interface to the WFS. Key components are labelled.}\label{fig:wfs_mech}
\end{figure}

The WFS will be located in the Coud\'{e} labs of the VLT Unit Telescopes. Fig.~\ref{fig:wfs_mech} shows an overview of the mechanical structure supporting the WFS optics and cryostat. The incoming light beam is shown in yellow, and the STS Unit is contained within the blue box. The WFS cryostat is mounted inside the tower.

Part of the optical assembly of the WFS will be operated at cryogenic temperature. The main cryostat is a simple LN$_2$ two-tank bath design based on those currently in use for other ESO instruments. However the detector characteristics, which are much improved at temperatures below that provided in the LN$_2$ cryostat, provide a strong incentive to cool the WFS to 40K. We are thus pursuing a development programme for a low-vibration pulse tube cooler, which will improve the WFS overall performance significantly.

\subsection{Adaptive optics real-time control}
\label{sec:wfs_rtc}
                                     
Compensation of atmospheric turbulence effects in astronomical images requires fast real-time measurement and wavefront correction. The WFS for GRAVITY will be operated at 500 Hz when a bright guide star is available (mK=7), for fainter guide stars (mK=10) this will be reduced to 200 Hz (see Section~\ref{sec:aoperf}).

The required computations for the wavefront sensing and subsequent commands to the MACAO deformable mirror are performed by the real time computer (RTC). A dedicated RTC application optimized for GRAVITY will be built on the newly developed SPARTA-Light platform, the low-cost implementation of ESO's Standard Platform for Adaptive Optics Real Time Applications (SPARTA)\cite{suarez_spie12}. 

Detailed AO performance simulations were performed for this optimization, examining the number of subapertures, operating frequency, centroiding window and algorithm, given different atmospheric conditions, WFS bandpasses, guide star magnitudes and WFS operational modes. Key results from these simulations are presented in Section~\ref{sec:aoperf}.

\subsection{Laboratory testing of the wavefront sensors}
\label{sec:wfs_testing}

Each of the 4 WFS units will be fully set up in the adaptive optics lab at the Max-Planck-Institut f\"{u}r Astronomie (MPIA) in Heidelberg for detailed functional and performance testing prior to delivery to Paranal and on-sky commissioning. A dedicated optical setup was designed to inject a VLT-like beam into the WFS under test. Paranal-like turbulence can be introduced into the path using a turbulence generator designed and built in-house.

A MACAO deformable mirror spare was provided on loan to MPIA by ESO, identical to the devices installed at VLT, in order to test the WFS fully in closed loop. This will allow us to perform full functional testing and optimization of the RTC and instrument control software, simulate realistic observations, and test calibration strategies.

\section{ADAPTIVE OPTICS PERFORMANCE}
\label{sec:aoperf}

Detailed performance simulations were performed to optimise the WFS design and operational parameters using end-to-end Monte Carlo simulation software YAO, v. 4.5.1., written by F. Rigaut\footnote{http://frigaut.github.com/yao/}. Methods and early results are presented in detail in Cl\'{e}net et al\cite{clenet_spie10}. The input atmosphere parameters, summarised in Table~\ref{tab:atmosphere}, are characteristic of conditions at Paranal; simulations covered ``nominal'' and ``pessimistic'' conditions (see table caption).

\begin{table}
	\centering
	\begin{tabular}{|c|c|c|c|}
		\hline
		Layer number & Altitude (km) & C$_{n}^{2}$ (\%) & Wind speed (m/s)\\
		\hline
		1 & 0 & $a \times 41$ & $a^{-3/5} \times 10$ \\
		2 & 0.3 & $a \times 16$ & $a^{-3/5} \times 10$ \\
		3 & 0.9 & $a \times 10$ & $a^{-3/5} \times 6.6$ \\
		4 & 1.8 & $b \times 9$ & $b^{-3/5} \times 12$ \\
		5 & 4.5 & $b \times 8$ & $b^{-3/5} \times 8$ \\
		6 & 7.1 & $b \times 5$ & $b^{-3/5} \times 34$ \\
		7 & 11 & $b \times 4.5$ & $b^{-3/5} \times 23$ \\
		8 & 12.8 & $b \times 3.5$ & $b^{-3/5} \times 22$ \\
		9 & 14.5 & $b \times 2$ & $b^{-3/5} \times 8$ \\
		10 & 16.5 & $b$ & $b^{-3/5} \times 10$ \\
		\hline
	\end{tabular}
	\smallskip
	\caption{Parameters of the input atmosphere for the YAO AO performance simulations. $a=1$ for a ``nominal'' C$_n^2$ profile, $a=0.7$ for the ``pessimistic'' case. $b=(1-0.67 \times a)/0.33$.}\label{tab:atmosphere}
\medskip
\end{table}

The parameter space explored with the simulations examined in particular:
\begin{itemize}
	\item{Number of subapertures;}
	\item{Number of pixels per subaperture;}
	\item{K-band magnitude of the reference source;}
	\item{Loop frequency;}
	\item{Loop gain;}
	\item{Subaperture flux threshold; and}
	\item{Seeing value.}
\end{itemize}

\begin{figure}
	\centering
	\includegraphics[width=12cm]{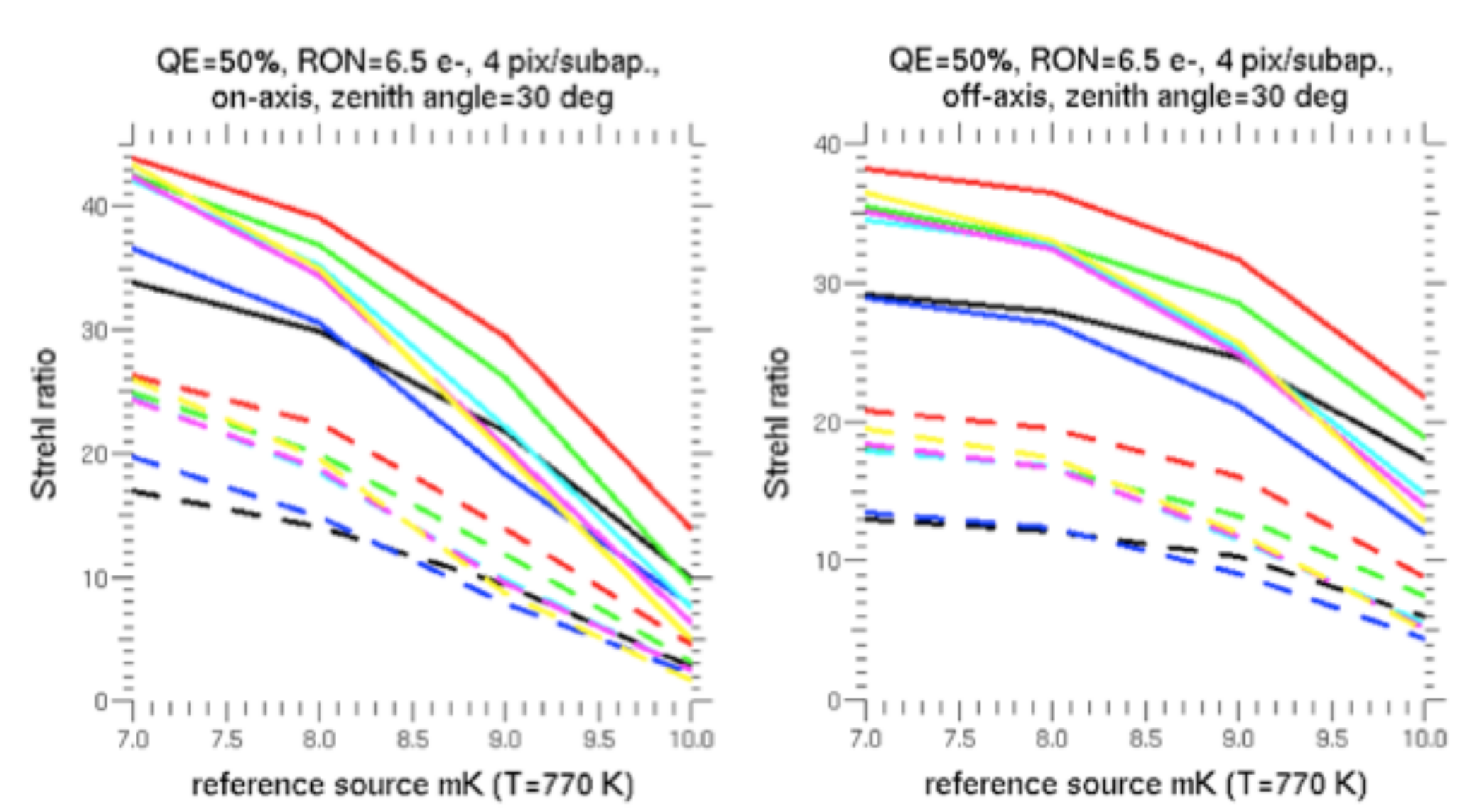}
	\caption{YAO simulations results for on-axis (left) and off-axis (right) operation of the GRAVITY WFS, valid after the DM. Solid lines assume nominal seeing values of 0.85'', dashed lines a more pessimistic 1.15''. Black, red, green, blue, cyan, magenta and yellow lines represent 8$\times$8, 9$\times$9, 10$\times$10, 11$\times$11, 12$\times$12, 13$\times$13 and 14$\times$14 subapertures, respectively. For each reference source, the Strehl ratio plotted uses the best value in the explored parameter space in frequency, gain and threshold. }\label{fig:aoperf_plot}
	\smallskip
\end{figure}

Key results from these simulations are shown in Fig.~\ref{fig:aoperf_plot}, which plot Strehl ratio against guide star K-band magnitude, for on- and off-axis operation (left and right plots, respectively). The plots assume a QE of 50\%, 6.5 e- read noise, centroiding performed over 4$\times$4 pixels per subaperture and a zenith angle of 30\degree. A 2-frame delay was introduced between the slope measurement and the command application to the DM. The Strehl ratios shown are valid after the DM only; additional sources of error, including non-commnon path errors, telescope optical image quality and vibration residuals, were incorporated for a full error budget analysis. The best overall performance is provided by the 9$\times$9 subapertures configuration, and this was chosen for the optical design and specification of the micro-lens array.

In off-axis operation the WFS receives 100\% of the light from the reference source, while in on-axis mode 50\% of the light of the target is directed to the beam combiner instrument. The resulting difference in transmission (15\% for on-axis, 32\% for off- axis) explains the improved results in off-axis mode at low flux. At high fluxes, the lower transmission of the on-axis compared to the off-axis mode is ``compensated'' by the lack of anisoplanatic error.

Study of the loop frequencies yielded an optimal frequency of 500 Hz for bright guide stars (mK=7). Performance with mK=10 guide stars requires lowering of the frequency to 200 Hz for best results.  Loop gain and pixel threshold values are varied according to guide star magnitude to optimize the overall correction bandwidth of the WFS.
                       
The residual AO errors were combined with additional sources of image degradation to compute a full error budget. The resulting wavefront errors are shown in Table~\ref{tab:aoperf} and compared with the corresponding requirement. In all cases the expected performance at 80K is satisfactory against the requirements, however given the parallel development of the detectors some uncertainty remains on the final detector performance. Additional cooling to 40K is expected to improve detector cosmetics, read noise and QE, which will significantly reduce the residual WFE.

\bigskip

\begin{table}
	\centering
	\begin{tabular}{|l|p{2.5cm}|p{2.5cm}|}
		\hline
		\textbf{AO Reference star} & \textbf{Required WFE (nm)} & \textbf{Estimated WFE (nm)}\\
		\hline
		mK = 7 (on-axis, z=30\degree) & 359 & 363 \\
		mK = 7 (off-axis, z=30\degree) & 412 & 393 \\
		mK = 10 (on-axis, z=30\degree) & 531 & 522 \\
	\hline
	\end{tabular}
	\smallskip
	\caption{Summary of AO simulation results compared with the requirement, expressed as residual wavefront error (WFE). Results assume a median seeing of 0.85'' at zenith and a QE of 50\%.}\label{tab:aoperf}
	\medskip
\end{table}

\section{SUMMARY}
\label{sec:summary}                         

The GRAVITY WFS development programme will enable adaptive optics wavefront sensing at IR wavelengths for the first time at the VLT Interferometer, enabling the use of IR-bright reference sources for the GRAVITY instrument. Detailed performance simulations show that the WFS with a Shack-Hartmann optical design scheme will provide 360 nm rms residual WFE for the case of a mK=7 AO guide star observed on-axis, in line with the required performance. The WFS will benefit greatly from the use of the new SELEX/ESO SAPHIRA eAPD detector arrays, which will for the first time allow frame rates of $\sim$500~Hz with low read noise (6.5 e- rms) compared with current state-of-the-art NIR detectors. Part of the instrument will be cooled to LN2 temperatures, with an upgrade path for pulse tube cooling to 40K under active development.

The WFS will be comprehensively tested in the adaptive optics lab at MPIA, where closed loop performance will be achieved with the aid of an atmospheric seeing generator and a MACAO deformable mirror identical to the units installed at VLT. The first GRAVITY WFS is expected to see first light at Paranal in 2014.

\bibliographystyle{spiebib}   

\end{document}